\documentclass[10pt,conference]{IEEEtran}
\IEEEoverridecommandlockouts
\usepackage{booktabs}
\usepackage{multirow} 
\usepackage{diagbox}
 
\usepackage{amsmath,amsthm,amssymb,amsfonts}
\usepackage{algorithm}
\usepackage{algorithmic}
\usepackage{listings}
\usepackage{amsthm}
\usepackage{graphicx}
\usepackage{subcaption}
\usepackage{tabularray}
\usepackage[dvipsnames]{xcolor}
\usepackage{caption}
\usepackage{float} 
\usepackage{tcolorbox}

\usepackage{xspace}
\usepackage{url}
\usepackage{hyperref}
\tcbuselibrary{breakable,skins}
\usepackage{tabularx,booktabs,array}
\usepackage{xcolor}

\usepackage{marvosym}
\newcommand{\framework}{\textsf{Malaika}\xspace}
\begin{document}

\title{\framework: Understanding Malware through Tri-Grounded Agentic Reasoning}

% \author{\IEEEauthorblockN{Xingzhi Qian}
% 	\IEEEauthorblockA{University College London\\
% 		someemail@somedomain.com}
% 	\and
% 	\IEEEauthorblockN{Homer Simpson}
% 	\IEEEauthorblockA{Twentieth Century Fox\\
% 		homer@thesimpsons.com}
% 	\and
% 	\IEEEauthorblockN{James Kirk\\ and Montgomery Scott}
% 	\IEEEauthorblockA{Starfleet Academy\\
% 		someemail@somedomain.com}}
\author{\IEEEauthorblockN{Xingzhi Qian,
        Xinran Zheng,
        Yiling He\textsuperscript{\Letter}
        and
        Lorenzo Cavallaro}
    \IEEEauthorblockA{University College London}
    \IEEEauthorblockA{\{xingzhi.qian.23, xinran.zheng.23, yiling-he, l.cavallaro\}@ucl.ac.uk}
    \thanks{\Letter~Corresponding Author.}
}

% \IEEEoverridecommandlockouts
% \makeatletter\def\@IEEEpubidpullup{6.5\baselineskip}\makeatother
% \IEEEpubid{\parbox{\columnwidth}{
% 		Network and Distributed System Security (NDSS) Symposium 2026\\
% 		23 - 27 February 2026 , San Diego, CA, USA\\
% 		ISBN 979-8-9919276-8-0\\  
% 		https://dx.doi.org/10.14722/ndss.2026.[23$|$24]xxxx\\
% 		www.ndss-symposium.org
% }
% \hspace{\columnsep}\makebox[\columnwidth]{}}

\maketitle

\begin{abstract} 
    Recent LLM-based systems have shown promising capabilities for security-focused code analysis, including vulnerability identification and reverse engineering. Malware understanding, however, poses a distinct challenge: analysts must reconstruct high-level malicious behaviors under partial observability from sparse, dispersed evidence intertwined with benign functionality. While static analysis can expose security-relevant signals, the central challenge is not merely identifying suspicious code, but determining whether the evidence sufficiently supports an auditable behavior-level conclusion. We formulate malware understanding as a grounded reasoning problem and argue that reliable behavior reconstruction requires three complementary forms of grounding. Domain grounding constrains how behavior hypotheses are generated and evaluated, semantics grounding localizes and connects supporting program evidence, and knowledge grounding supports behavioral attribution through externally verifiable threat knowledge. To study this hypothesis, we present \framework, a tri-grounded multi-agent framework that operationalizes the three grounding mechanisms through analyst-inspired reasoning, tool-mediated evidence localization, explicit review, and retrieval-based behavioral attribution. We instantiate \framework for Android malware analysis and evaluate it on malware-understanding tasks with analyst-validated behavior-level annotations. Our results show that \framework improves analysis quality over prior LLM-based malware-analysis frameworks and demonstrate that reliability depends not only on model capability but also on the structure of the reasoning process itself. In particular, comparisons against both malware-analysis systems and frontier agentic frameworks show that behavioral attribution is strongly influenced by the surrounding reasoning process, with grounding-aware reasoning producing substantially more precise and auditable conclusions. More importantly, ablation studies support the grounding hypothesis: domain grounding, semantics grounding, and knowledge grounding address complementary failure modes by contributing distinct capabilities for hypothesis generation, evidence localization, and behavioral attribution, respectively. These findings suggest that grounding-aware reasoning provides a principled foundation for reliable malware understanding and, more broadly, for evidence-grounded software analysis.

\end{abstract}

\IEEEpeerreviewmaketitle
\section{Introduction}

Malware analysis remains a central problem in software security. In practice, a binary prediction of whether an application is malicious is rarely enough for analysts. They need to understand what the program does, how the behavior is triggered, and which concrete code-level evidence supports each conclusion. Reliable malware understanding therefore requires more than detecting suspicious signals. It requires linking low-level program facts to high-level behavioral interpretations in a way that is both auditable and grounded in evidence.

Existing malware detection systems have made substantial progress in scalable classification. Conventional learning-based detectors abstract applications into classification-oriented representations, from manually engineered features~\cite{arp2014drebin, deepdrebin2017} to learned representations over richer program structures and temporal signals~\cite{he2022msdroid, cade2021, continuous2023, zheng2025tif}. These methods are useful as front-line detectors, but their outputs are usually malware labels or feature-level explanations~\cite{he2023finer}. Such outputs provide limited support for the downstream task of reconstructing concrete malicious behavior and checking whether each claim is supported by evidence. They can often indicate that an application is suspicious, but they do not reliably explain how the application implements malicious logic.

LLMs offer a promising way to narrow this gap. Recent LLM-based systems have shown strong capabilities in security-focused code analysis, including vulnerability detection, binary reverse engineering, and malware analysis~\cite{ullah2024llms,sun2024llm4vuln,zheng2025maleval}. Their ability to produce natural-language explanations from code makes them useful for security workflows that require human-readable reasoning. However, malicious-code understanding remains difficult because the task is not simply to summarize a local code snippet. A system must examine partial and fragmented code snippets, form possible behavior hypotheses, and synthesize a final analysis from concrete evidence. Prior LLM-based malware-analysis pipelines, including hierarchical prompting~\cite{walton2024malparse} and slicing-based approaches~\cite{qian2025lamd}, have demonstrated the feasibility of applying LLMs to malware analysis, but they still tend to rely on limited fixed analysis pipelines, where the analysis scope is bounded by restricted sources. They also do not explicitly control how hypotheses are generated, checked, and tied back to evidence.

Frontier coding agents, such as Codex~\cite{chen2021codex} and Claude Code~\cite{anthropic2026claudecodesecurity}, have shown strong performance on code-related and security-critical tasks, especially vulnerability discovery and validation~\cite{openai2026codexcybersafety,carlini2026mythos}. However, their behavior is still shaped by the surrounding prompt and skill design. 
Recent evidence shows that, even when the data and parsing are fixed, changing only the prompting can substantially alter model behavior, making prompt sensitivity a first-class system property~\cite{camarato2026promptaudit, ferrari2026llmobfuscation}. 
This raises a more specific question than whether frontier models can reason about security code reliably. When moving from vulnerability discovery to malicious-code understanding, what kind of harness is needed to make reasoning reliable, evidence-grounded, and deployable under realistic cost and data-governance constraints?
% Recent evidence shows that, even when the dataset, decoding, and parsing are fixed, changing only the prompting strategy can substantially alter model behavior, making prompt sensitivity a first-class system property~\cite{camarato2026promptaudit}. 
% This raises a more specific question than whether frontier models can reason about security code reliably. When moving from vulnerability discovery to malicious-code understanding, what kind of harness is needed to make reasoning reliable, evidence-grounded, and deployable under realistic cost and data-governance constraints?
% This observation motivates a broader question: should reliability emerge primarily from prompt engineering, or from explicit reasoning constraints? We argue for the latter. Rather than relying on task-specific prompt optimization, \framework embeds analyst expertise into a structured reasoning workflow through domain grounding, semantics grounding, and knowledge grounding. This shifts the focus from prompt design toward reproducible and inspectable reasoning mechanisms.

The distinction matters because vulnerability discovery and malware understanding have different reasoning goals. General-purpose coding agents are largely designed for capability and coverage. They provide file-system tools, broad repository access, and large reasoning budgets so that a strong model can inspect more code and find relevant defects~\cite{yang2024sweagent, wang2025openhands}. This design fits many vulnerability-discovery settings, where a central challenge is locating the code region that contains a defect~\cite{zhangvulteller, zheng2026veritas}. Malware understanding, however, is a reasoning problem under partial observability. A common failure mode is not that the system never inspects suspicious code, but that it turns partial evidence into an overconfident behavioral claim. This risk is especially problematic because the same security-sensitive mechanisms can support both benign and malicious functionality depending on the triggering context and surrounding program semantics \cite{appcontext2015}. The main challenge is therefore verification and attribution: whether a behavior hypothesis can be linked to concrete program semantics and sufficient supporting evidence.

We argue that reliable malicious-code understanding requires three complementary forms of grounding. The system needs domain grounding to structure analysis in a manner consistent with professional security practice. It also needs semantics grounding so that it knows where to inspect by using related analysis tools and understanding the program structures and contexts, rather than only generic file browsing or keyword search. Finally, it needs knowledge grounding, where observed behaviors are connected to externally verifiable threat knowledge without treating information gained during model training as proof. These forms of grounding are not just engineering modules. They are mechanisms for turning a capable model into a more reliable malware analyst.

To study this design point, we propose a general tri-grounded multi-agent framework for malware understanding. We instantiate the framework in the Android ecosystem not because the proposed grounding principles are Android-specific, but because Android provides a realistic setting where behavior reconstruction is challenging, behavior-level annotations are available, and mature threat taxonomies enable systematic evaluation.
% , and use Android malware analysis as a concrete instantiation due to its combination of platform-specific execution semantics, dispersed code-level evidence, and mature behavior taxonomies. 
We realize this Android instantiation as \framework, where the grounding principles are realized through an analyst-inspired workflow consisting of manifest analysis, path exploration, evidence review, and final report synthesis. Domain grounding structures the reasoning process as an iterative interaction between Explorer agents, which propose and refine behavior hypotheses, and Reviewer agents, which assess whether the supporting evidence is sufficient and logically consistent. Semantics grounding provides static-analysis tools for localizing suspicious regions and retrieving function summaries, class-level context, caller context, and execution-relevant program facts. Knowledge grounding retrieves relevant MITRE ATT\&CK mobile techniques only after concrete behaviors have been observed, so that threat knowledge verifies and contextualizes evidence-backed claims rather than replacing code-level evidence.

We evaluate \framework on the full MalEval benchmark~\cite{zheng2025maleval} and a targeted 20-sample Mobile ATT\&CK attribution audit. \framework improves report-level malware-understanding quality over prior LLM-based malware analysis approaches, showing that grounding-aware reasoning improves behavior-level reports beyond fixed, single-pass pipelines. Compared with frontier coding agents, \framework provides a more precision-oriented and lower-cost operating point. Ablation results further show that reliability is strongly shaped by the harness rather than by the base model alone. When it is wrapped in the grounded harness of \framework, the model produces analyses that are more closely tied to evidence. These findings suggest that scaling model capability and context length alone is insufficient for reliable malware reasoning. The harness around the model is a central part of system reliability.

% This paper formulates malicious-code understanding as a grounded reasoning problem under partial observability, where the central challenge is not only to explore code but also to verify whether behavioral hypotheses are sufficiently supported by concrete evidence. It identifies three complementary requirements for reliable malware reasoning: domain grounding for analyst-inspired hypothesis review, semantics grounding for execution-relevant evidence localization, and knowledge grounding for externally verifiable behavioral attribution. It presents \framework, a tri-grounded multi-agent framework that combines iterative analyst-style reasoning, Android-aware evidence localization, explicit review, and MITRE ATT\&CK retrieval to produce auditable malicious-behavior reports. It also empirically studies the capability--reliability--cost trade-offs among \framework, prior LLM-based malware-analysis systems, and frontier coding agents, showing that a grounded harness can achieve strong precision-oriented malware understanding with an open-weight model under realistic deployment constraints.

In summary, this paper makes the following contributions:

\begin{itemize}
    \item We formulate malicious-code understanding as a grounded reasoning problem under partial observability, where the central challenge is not only to explore code but also to verify whether behavioral hypotheses are sufficiently supported by concrete evidence.
    
    \item We identify three complementary requirements for reliable malware reasoning: domain grounding for analyst-inspired hypothesis review, semantics grounding for execution-relevant evidence localization, and knowledge grounding for externally verifiable behavioral attribution.
    
    \item We present \framework, a tri-grounded multi-agent framework that combines iterative analyst-style reasoning, Android-aware evidence localization, explicit review, and MITRE ATT\&CK retrieval to produce auditable malicious-behavior reports.
    
    \item We empirically study the capability--reliability--cost trade-offs among \framework, prior LLM-based malware-analysis systems, and frontier coding agents, showing that a grounded harness can achieve strong precision-oriented malware understanding with an open-weight model under realistic deployment constraints.
\end{itemize}

\section{Motivation}

% Static analysis is indispensable for malware understanding. It can identify sensitive API calls, extract suspicious strings, and build call-graph relations. These artifacts provide the raw material for reasoning, but they are not behavior-level explanations. Security-relevant logic is often distributed across framework callbacks, asynchronous dispatch, and background services. A raw call-graph view may therefore miss the semantic relation that makes a behavior malicious, while isolated features such as permissions or APIs may encourage overconfident conclusions.

\begin{figure}[t]
\centering
\includegraphics[width=\columnwidth]{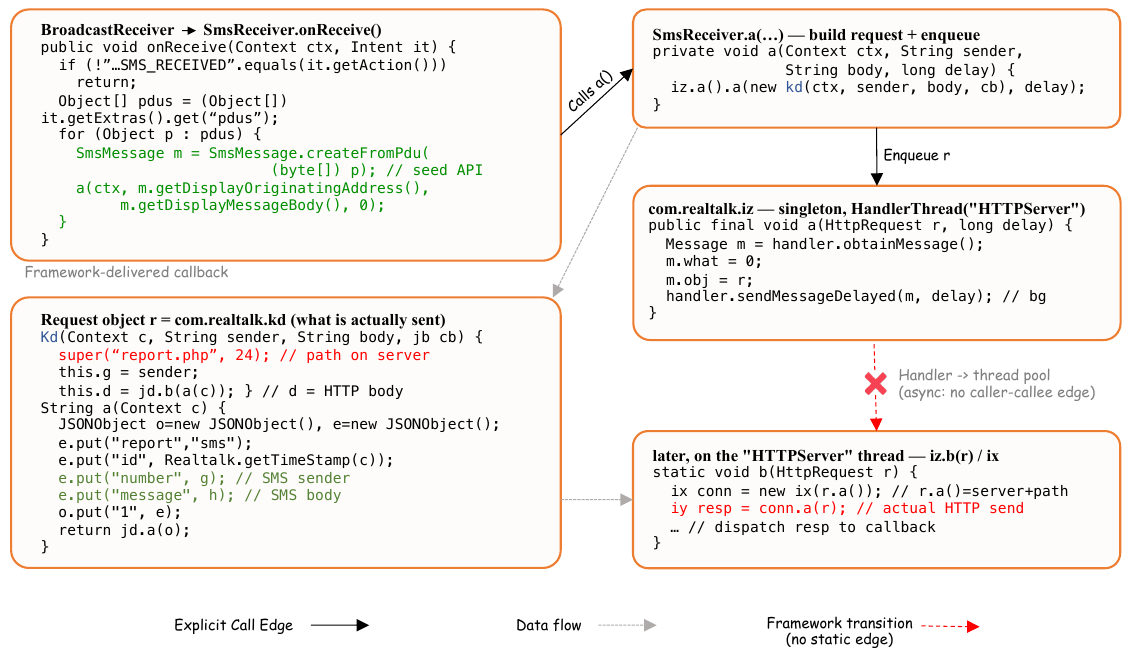}
% \caption{A motivation example of an SMS-handling path.}
\caption{Motivating example of fragmented behavior reconstruction. An SMS receiver extracts the sender and message body, packages them into a request, and enqueues the request for background execution. The eventual HTTP transmission occurs through an asynchronous framework-mediated transition, separating the SMS source from the network sink.}
\label{fig:motivation}
\end{figure}

Static analysis is indispensable for malware understanding because it exposes sensitive API calls, suspicious strings, permissions, entry points, and call-graph relations. However, isolated static facts can easily lead to incomplete or overconfident conclusions, as malware logic is often distributed across components. Figure~\ref{fig:motivation} illustrates this gap using a real sample. 
\texttt{SmsReceiver.onReceive} invokes \texttt{SmsMessage.createFromPdu} to parse incoming SMS messages, 
% The method extracts the sender and message body, packages them into a request object, and enqueues the request on a background handler thread. 
% The actual network transmission occurs later, when the queued request is dispatched to an HTTP connection wrapper.
while the SMS source and the actual network sink are connected through framework delivery and asynchronous state, which cannot be captured through a direct application-level call chain.
This highlights why static facts must be interpreted before they can support behavior-level claims. A fixed-depth call-graph traversal may stop at the enqueue operation and fail to connect the captured SMS content to the later network sink. In addition, visible indicators such as permissions or sensitive APIs indicate potential attack surface, but do not by themselves prove malicious abuse. Reliable malware understanding therefore requires reconstructing how dispersed program fragments interact and accepting a behavioral claim only when it is supported by concrete execution-relevant evidence.

\section{Methodology}
\begin{figure*}[t]
    \centering
    \setlength{\abovecaptionskip}{0cm}
    \includegraphics[width=0.90\linewidth]{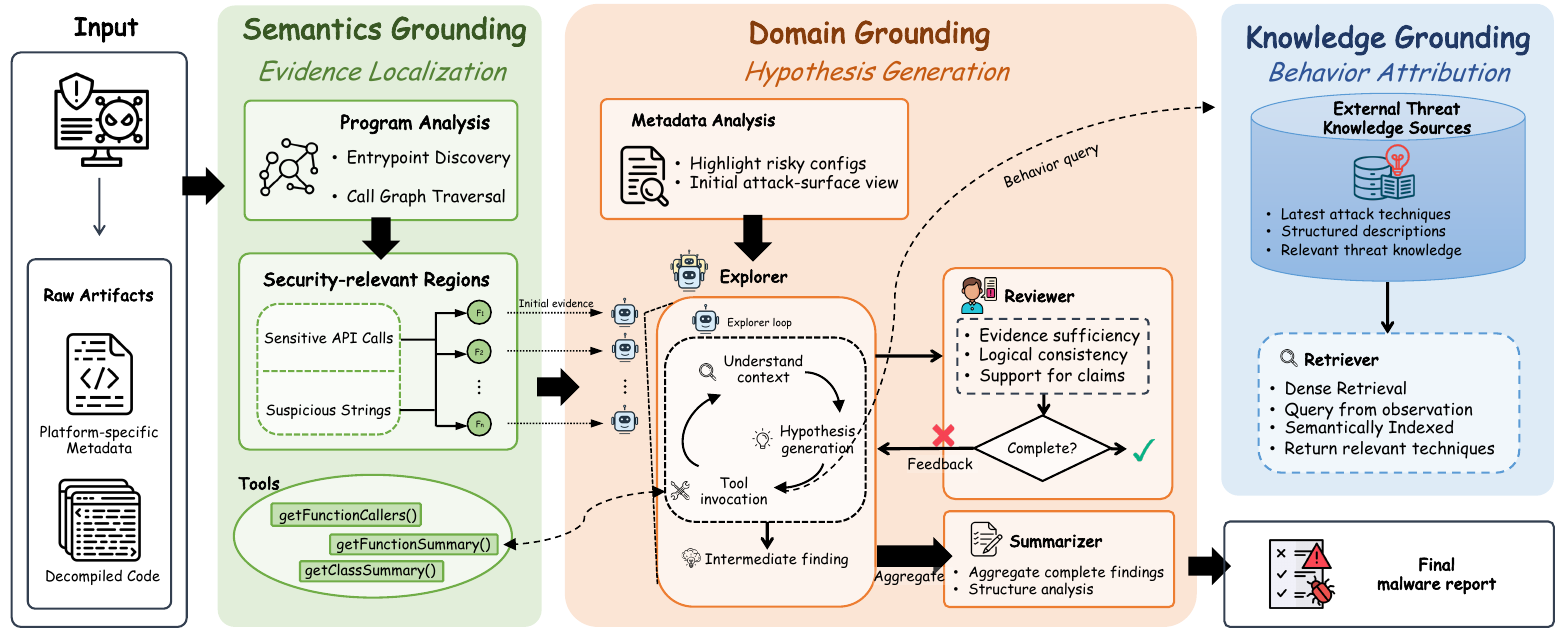}
    \caption{Architecture of \framework. 
 % \textit{Semantics grounding} extracts program facts from the APK and localizes security-relevant regions using entry-point discovery, call-graph traversal, sensitive API calls, and suspicious strings. \textit{Domain grounding} structures reasoning as a role-specialized multi-agent workflow: the \textit{Manifest Analyzer} produces an attack-surface prior, and each candidate seed is analyzed by a dedicated \textit{Explorer–Reviewer} thread. The \textit{Explorer} invokes analysis tools to expand program context and generates behavior hypotheses, while the paired \textit{Reviewer} verifies evidence sufficiency and logical consistency before accepting the result. \textit{Knowledge grounding} provides on-demand MITRE ATT\&CK retrieval based only on observed behavior, supporting attribution without replacing code-level evidence. Verified results are aggregated by the \textit{Summarizer}, which produces the final evidence-supported malware report.
 \framework analyzes a program by combining three forms of grounding: 
 \textit{Domain grounding} structures reasoning as a role-specialized multi-agent workflow,
 \textit{Semantics grounding} extracts program facts and localizes security-relevant regions, and
 \textit{Knowledge grounding} provides on-demand threat knowledge based on observed behavior.
 % \textit{domain grounding} organizes the reasoning process into analyst-inspired roles, \textit{semantics grounding} localizes and exposes program-level evidence, and \textit{knowledge grounding} supports behavior attribution with retrievable threat knowledge.
    }
    \label{fig:arch}
\end{figure*}

\subsection{Problem Formulation}

We formulate malware analysis as a grounded reasoning task. Given a program, the goal is not simply to output a malware label, but to iteratively build and verify hypotheses about potentially malicious behavior, starting from suspicious parts.
% This formulation is motivated by the nature of Android applications. Behaviorally relevant evidence is often scattered across multiple callbacks, components, and asynchronous execution paths. Moreover, many important relationships are implicit and mediated by the Android framework, rather than directly visible from explicit caller-callee relationships. As a result, reasoning over the static call graph alone, or over a fixed static slice, can easily miss crucial evidence, thus leading to unsupported conclusions.
Malware analysis can therefore be represented as a reasoning trajectory:
\[
\mathcal{R}(a) = \{(h_t, C_t, E_t)\}_{t=0}^{T},
\]
where $a$ is the input application, $h_t$ is the current behavior hypothesis at step $t$, $C_t$ is the accumulated analysis context, and $E_t$ is the supporting evidence collected so far. The system starts from suspicious program regions and repeatedly refines its understanding until the evidence is sufficient for a reliable conclusion.

% At each step, the reasoning process must answer three questions: where to look next, whether the current evidence is sufficient, and whether the current hypothesis should be accepted or revised. Hypotheses are not accepted simply because they appear plausible; instead, they must be supported by sufficient evidence and remain consistent with the observed program semantics. 
% We represent this decision as
% \[
% r_t =
% \begin{cases}
% \texttt{accept}, & \text{if the evidence is sufficient and consistent,} \\
% \texttt{revise}, & \text{otherwise.}
% \end{cases}
% \]
% Only accepted reasoning states are used to generate the final analysis decision.

This formulation emphasizes that the core challenge of malware analysis lies in reasoning under partial observability. A reliable system must localize relevant evidence, iteratively refine behavior hypotheses, and verify conclusions before presenting them to analysts. Motivated by this view, we believe three complementary forms of grounding are necessary for the harness: domain grounding, which structures reasoning as an analyst-inspired workflow; semantics grounding, which provides access to execution-relevant program semantics; and knowledge grounding, which aligns inferred behaviors with external threat intelligence.

\subsection{Framework Overview}

\framework is a tri-grounded multi-agent framework for malware understanding. 
As shown in Figure~\ref{fig:arch}, \framework analyzes a program by combining three complementary forms of grounding: \textit{Domain grounding}, \textit{Semantics grounding}, and \textit{Knowledge grounding}. 
% The framework is designed to produce not only a malware verdict, but also an auditable explanation grounded in observed program behavior.
% \framework is a tri-grounded multi-agent framework for malware understanding. 
The overall workflow consists of three stages. First, \framework preprocesses the program to extract raw artifacts and performs program analysis to identify entry points, call-graph relations, sensitive API invocations, and suspicious strings. These artifacts are used to construct a set of candidate analysis targets, marking specific functions as starting points for detailed behavioral analysis.

Second, \framework analyzes candidate targets through localized Explorer--Reviewer threads. For each candidate analysis target, an Explorer investigates the surrounding program context and constructs a behavior hypothesis supported by code-level observations. The Explorer can expand the analysis scope when the current context is insufficient, such as by inspecting callers, related functions, or class-level behavior. This allows \framework to focus on security-relevant regions while avoiding exhaustive whole-program inspection. Each Explorer output is audited by a paired Reviewer. The Reviewer checks whether the conclusion is supported by sufficient evidence, whether the reasoning is logically consistent with the observed program semantics, and whether any reported behavior attribution is properly grounded. If the evidence is incomplete or contradictory, the Reviewer returns feedback and requires the Explorer to continue analysis. Only accepted localized findings are propagated to the final stage.

Finally, the Summarizer aggregates accepted findings into a structured malware analysis report. The report includes the application-level verdict, observed behaviors, atomic evidence, and confirmed behavior mappings when applicable. This staged design separates evidence collection, verification, and final synthesis, reducing the risk that plausible but unsupported intermediate hypotheses are propagated into the final output.

\subsection{Domain Grounding}

Domain grounding constrains how malware reasoning is organized. Instead of allowing a single agent to inspect code, conduct analysis, and produce a final result, \framework follows an analyst-inspired workflow through four distinct roles.
The Manifest Analyzer summarizes the application's coarse-grained attack surface and highlights risky parts. The Explorer investigates localized analysis targets and constructs behavior hypotheses from observed program context. The Reviewer audits each Explorer output for evidence sufficiency and logical consistency. Finally, the Summarizer aggregates accepted findings into the final report. These roles are not merely implementation modules; they define constraints on what each stage is allowed to infer and propagate.
% This design is motivated by the ambiguity of local security signals in Android applications. Sensitive permissions, suspicious strings, or security-relevant API calls can indicate potential risk, but they are often inconclusive in isolation. A permission may support benign functionality, a suspicious string may be unused, and a sensitive API call may not lead to data leakage or unauthorized behavior. Reliable analysis therefore requires a disciplined process that separates hypothesis generation from evidence validation.

The Explorer--Reviewer loop is the central mechanism of domain grounding. The Explorer may propose tentative behavior hypotheses, but these hypotheses are not directly accepted. The Reviewer must first determine whether the claimed behavior is supported by concrete code-level evidence, whether the reasoning correctly connects relevant operations, and whether any external behavior attribution is tied to observed evidence. If the evidence is incomplete or contradictory, the Reviewer returns feedback and the Explorer continues analysis.
This separation is especially important for malware understanding because local code snippets can appear suspicious without broader execution context. For example, a permission-related API call may be benign if used for user-visible functionality, while an ordinary utility function may become security-relevant when connected to hidden triggers or external transmission. By separating hypothesis generation from evidence verification, \framework encourages conservative and evidence-grounded conclusions.

We further realize domain grounding through carefully designed prompts. Instead of relying on flexible user instructions, which may lead to inconsistent exploration strategies and results, our prompts specify an expert-inspired analysis procedure. 
% The Explorer is instructed to form and refine hypotheses from static evidence, expand context when support is insufficient, and avoid over-attribution. The Reviewer is prompted to check evidence sufficiency, logical consistency, and possible benign explanations. 
Few-shot examples are included to demonstrate how program facts should be converted into behavior-level findings.
% They illustrate the expected evidence format, the difference between suspicious APIs and confirmed behaviors, and the support required for ATT\&CK attribution. 
In this way, prompting operationalizes domain grounding by steering the agent toward evidence-centric malware analysis rather than generic code summarization.

\subsection{Semantics Grounding}
Semantics grounding provides the program-level basis for malware reasoning. Rather than asking an LLM to reason directly over an entire APK or raw decompiled code, \framework exposes selected program artifacts through static-analysis-based localization and tool-mediated context retrieval.

\textbf{Artifact Extraction and Target Selection.} Given an input APK, \framework extracts manifest information, decompiled source code, and string constants, and identifies entry points, call-graph relations, and permission-related sensitive API calls. The manifest provides component declarations, permissions, intent filters, and other configuration-level signals. Decompiled code and string constants provide function-level and class-level context. Call-graph relations provide explicit caller-callee structure for navigating program context.

\framework identifies entry points from both manifest-declared Android components and user-defined methods overriding Android framework APIs, following prior Android static-analysis practice~\cite{zhang2014droidsift, zheng2025maleval}.
Starting from the identified entry points, \framework uses the constructed call graph to identify candidate analysis targets within statically reachable regions. A method is selected as a candidate target if it invokes permission-related sensitive APIs or contains suspicious security-relevant strings.
% Sensitive API calls indicate operations involving protected resources or privileged functionality, while suspicious strings may reveal network endpoints, commands, encoded payloads, file paths, or other security-relevant indicators.
For each candidate, \framework records the associated sensitive APIs and strings as initial leads for subsequent reasoning. These candidates serve only to prioritize exploration: they are not treated as complete execution traces or sufficient evidence of maliciousness, since actual behaviors may not be fully captured by explicit call-graph edges.
% Candidate analysis targets are used only as starting points for detailed reasoning, and their ranking reflects only the density of security-relevant indicators rather than any judgment of intent. They are not treated as sufficient evidence of maliciousness. This distinction is important because many benign applications invoke sensitive APIs or contain security-relevant strings for legitimate purposes. The maliciousness of a target can only be determined after the following context exploration stage. 
To prioritize analysis, \framework ranks candidate targets by the total number of associated sensitive APIs and suspicious strings, dispatching higher-ranked targets in earlier batches.
% This ranking determines the order in which targets are fed into the analysis stage, and under the parallel execution model it ensures that the most promising leads occupy available analysis slots first, improving the likelihood of surfacing malicious behavior early within a bounded analysis budget.

\textbf{Tool-Mediated Context Retrieval.} To support grounded reasoning while controlling context-window usage, \framework exposes compact static-analysis views to the Explorer through a set of tools: \texttt{getFunctionSummary}, which generates a function-level summary from decompiled code; \texttt{getFunctionCallers}, which returns the function callers based on the call graph; and \texttt{getClassSummary}, which retrieves class-level context when function-level evidence is insufficient. The Explorer invokes these tools adaptively, expanding the analysis context until the behavior is sufficiently supported.
% \begin{itemize}
%     \item \texttt{getFunctionSummary} retrieves a compact summary of a target function, including its main operations, invoked APIs, accessed data, and relevant control-flow or data-flow observations. It helps the Explorer reason about long or complex functions without loading excessive code context.

%     \item \texttt{getFunctionCallers} retrieves caller functions of the current function. It supports upward context expansion, allowing the Explorer to determine how an operation is triggered, and whether it is reachable from framework-dispatched callbacks.

%     \item \texttt{getClassSummary} retrieves class-level context when function-level evidence is insufficient. It is useful when behavior is distributed across fields, helper methods, inner classes, or lifecycle methods in the same class.
% \end{itemize}
% The Explorer invokes these tools adaptively. It expands context when current evidence is insufficient, and stops once the behavior is sufficiently supported for review. Tool outputs are cached to avoid repeated analysis and reduce redundant LLM calls.

% \textbf{Handling Framework-Mediated Android Execution} 
% Android applications are event-driven and framework-mediated, so
Behaviorally relevant relations are not always represented as explicit caller-callee edges in the static call graph. 
% Lifecycle callbacks may be invoked by the Android framework, broadcast receivers may be triggered by system events, services may be scheduled through \texttt{PendingIntent} or \texttt{AlarmManager}, and background tasks may be dispatched through threads, executors, or runnable objects.
In such cases, direct traversal over call-graph edges may miss important behavioral context or fail to explain how a suspicious operation is triggered. Different call-graph construction algorithms also make different trade-offs between precision and soundness~\cite{callgraphsoundness}. \framework instead leverages LLMs to understand the current context and decide the next step based on the existing analysis. This context-adaptive exploration helps recover behaviorally relevant context that may not be obvious from a single explicit call-chain view. As a result, semantics grounding enables \framework to reason over localized program behavior while remaining anchored to concrete static artifacts.

\subsection{Knowledge Grounding}
Knowledge grounding supports the attribution of observed behaviors to external threat knowledge. While semantics grounding provides code-level evidence, malware analysis also requires standardized interpretation of behaviors, such as mapping observed actions to known adversarial techniques. Relying only on an LLM's parametric knowledge can lead to outdated or unsupported technique mappings. \framework therefore uses an external MITRE ATT\&CK knowledge base to ground behavior attribution in retrievable and inspectable sources. This design allows \framework to accommodate the evolving nature of the knowledge base without modifying the underlying models or agents. When a new ATT\&CK version is released, the corresponding documents can be updated and the retrieval index can be rebuilt. Unlike fine-tuning, which requires periodically retraining as threat knowledge evolves, and in-context learning, which is constrained by context window, retrieval provides a lightweight way to maintain up-to-date threat knowledge.
% A MITRE ATT\&CK technique is reported only when the underlying behavior has already been observed in the analyzed code. Retrieved knowledge provides attribution support, not proof of maliciousness.
During exploration, the Explorer invokes the retrieval tool after identifying a concrete behavior that appears security-relevant. The tool returns candidate ATT\&CK techniques with their identifiers and descriptions. The Explorer may then include a technique mapping only if the retrieved technique matches the observed behavior. The Reviewer further checks attribution discipline. If an Explorer reports an ATT\&CK technique without retrieved support, or if the retrieved technique is not connected to concrete observed behavior, the Reviewer marks the analysis as incomplete. The Summarizer is also restricted to aggregating only the ATT\&CK mappings accepted during localized analysis; it must not introduce new technique IDs during final report generation.

This design improves auditability in two ways. First, technique mappings become traceable to an external knowledge base rather than relying solely on the LLM's internal knowledge. Second, attribution remains grounded in observed program behavior rather than in superficial similarity between malware descriptions and technique names. Knowledge grounding therefore complements semantics grounding: code evidence establishes what the application does, while retrieved threat knowledge helps describe how that behavior aligns with standardized adversarial techniques.

\section{Evaluation}
To evaluate the effectiveness of \framework, we conduct experiments to address the following research questions:

\begin{itemize}
    \item \textbf{RQ1:} How effective is \framework for malware understanding compared with existing LLM-based and frontier agentic systems?
    \item \textbf{RQ2:} To what extent are \textsf{Malaika}'s tri-grounding design choices necessary for achieving high-quality malware reasoning?
    \item \textbf{RQ3:} Does \framework produce reliable malware analysis with the proposed grounding mechanisms?
\end{itemize}

\subsection{Evaluation Setup}
\subsubsection{Dataset}
Android serves as an experimentally convenient setting for studying malware understanding because analyst-validated behavior annotations, platform semantics, and standardized behavior taxonomies are available. We evaluate \framework on the MalEval benchmark dataset~\cite{zheng2025maleval}, which contains 255 Android applications, including 230 malware samples across six representative malware categories and 25 benign applications. Unlike datasets designed primarily for malware classification~\cite{chow2025hypercube}, MalEval provides analyst-written reports and manually verified behavioral annotations, making it suitable for evaluating grounded behavioral reasoning and explanation quality. Although MalEval is relatively small, it is appropriate for our setting because behavior-level malware understanding requires fine-grained reports and manually verified behavioral annotations, which are substantially more expensive to construct than binary labels. We therefore use MalEval to evaluate whether systems can produce evidence-supported behavioral explanations, while discussing the generalization limits of this benchmark in Section~\ref{sec:threats}.

% First, in practice, analyzing every single program is expensive on both time-wise and cost-wise. Therefore, we envision \framework as a component in a larger analysis pipeline rather than a front-line classifier applied to every unknown sample. To provide an appropriate scenario, samples that are rejected or predicted with high uncertainty by learning-based models~\cite{cade2021,he2022msdroid,continuous2023, zheng2025tif} can be further analyzed by \framework to provide reliable and comprehensive analysis result, therefore automating the entire pipeline through continuously refining the model. 
% Second, our task is behavioral understanding rather than binary malware detection. This requires fine-grained ground truth over concrete behaviors and supporting evidence, which is substantially more expensive to construct and verify than binary classification.

% To assess generalization beyond the benchmark distribution, we further include a small set of recent Android malware samples from 2026 collected from public security reports. These samples are not covered by existing benchmarks and are used for out-of-distribution (OOD) analysis.
\subsubsection{Implementation}
We implement \framework using LangGraph~\cite{langgraph} for agent orchestration and Androguard~\cite{androguard} for static analysis. For knowledge grounding, we construct an Android-focused MITRE ATT\&CK Mobile index from the \texttt{mobile-attack-18.0} STIX bundle. We normalize relevant STIX attack-pattern objects, embed them with \texttt{text-embedding-3-large}, and store them in a FAISS index~\cite{faiss} for dense retrieval. For each behavior query, the retriever returns the top-10 relevant techniques.
\framework processes ranked candidates in batches of up to five parallel Explorer--Reviewer threads. Once five Reviewer-accepted malicious findings are obtained, it cancels remaining analyses and summarizes the verified findings; both limits are fixed before evaluation rather than tuned on benchmark performance.

\subsubsection{Baseline and Model Selection}
We evaluate against both state-of-the-art LLM-based Android malware detectors, LAMD~\cite{qian2025lamd} and MalEval~\cite{zheng2025maleval}, and two representative frontier agentic systems, Codex~\cite{openai2026codexcybersafety} and Claude Code~\cite{anthropic2026claudecodesecurity}, to provide a comprehensive comparison. LAMD performs program slicing and context-driven analysis around sensitive APIs, whereas MalEval represents an unguided single-pass LLM pipeline.

% \begin{itemize}
%     \item \textbf{LAMD~\cite{qian2025lamd}}: A context-driven LLM malware detection framework based on static slicing around sensitive APIs.
%     \item \textbf{MalEval~\cite{zheng2025maleval}}: The default LLM analysis pipeline from MalEval, representing unguided, single-pass reasoning without agentic workflow and RAG.
%     \item \textbf{Claude Code} and \textbf{Codex}: Representative frontier coding agents with high cybersecurity capability.
% \end{itemize}

% \subsubsection{Model Selection} 
Our goal is not to benchmark LLM capability in isolation, but to evaluate whether \framework can improve malware reasoning under modern model settings. 
% This task requires long-context code reasoning and reliable multi-step tool use, which are difficult to satisfy with small or less capable models. 
Using strong models therefore helps reduce failures caused by limited base-model capacity. We use \textbf{DeepSeek-V3.2}~\cite{liu2024deepseek} as the primary back-end model, which is one of the frontier models designed for agentic tasks, with temperature = 0. For frontier agentic baselines, Codex is evaluated with GPT-5.4 using medium reasoning effort, while Claude Code is evaluated with Opus-4.7 \textit{xhigh} reasoning. To measure the effect of domain-specific skill augmentation, we additionally evaluate Claude Code equipped with an Android-malware analysis skill.\footnote{\texttt{analyzing-android-malware-with-apktool}, from the repository: \url{https://github.com/mukul975/Anthropic-Cybersecurity-Skills}.} To further isolate harness effects from base-model capability, we instantiate Codex and Claude Code harnesses with DeepSeek-V3.2. Finally, we evaluate \framework with GPT-5.4 to assess whether its grounding mechanisms generalize well across other strong back-end models.

\subsubsection{Metrics}
We evaluate malware understanding from three complementary perspectives: report-level analysis quality, structured behavior attribution, and operational reliability. 
% Report-level metrics measure whether the generated report reaches the correct malware-level conclusion and captures meaningful behavioral explanations. Structured attribution metrics measure whether the system can map observed behaviors to standardized ATT\&CK techniques. Reliability and cost metrics measure whether the analysis is stable and reliable in real practice.

\textbf{Report-level Malware Understanding.} Following MalEval~\cite{zheng2025maleval}, we report five metrics: 
% binary F1 (B-F1), report quality (RQ), false positive correction rate (FPCR), true positive maintenance rate (TPMR), and category-level F1 ($F1_c$). 
Binary F1 (B-F1) evaluates the consistency between generated and ground-truth behaviors. Report quality (RQ) measures the overall quality of the generated malware report using llm-as-a-judge. False positive correction rate (FPCR) measures the system's ability to avoid falsely labeling malicious applications as benign, while true positive maintenance rate (TPMR) measures whether malware samples are correctly identified. Category-level F1 ($F1_c$) evaluates whether the system correctly distinguishes behaviors from different malware categories. We additionally report the failure rate for each malware category, computed as the proportion of misclassified samples in that category.
% In addition, we report category-level failure rates to show where systems struggle across different malware types. 
% Higher values are better for B-F1, RQ, FPCR, TPMR, and $F1_c$, while lower values are better for category-level failure rates.
\begin{table*}[t]
\caption{Report-level malware understanding results on MalEval dataset. 
Higher is better for B-F1, RQ, FPCR, TPMR, and $F1_c$; lower is better for category-level failure rates.
}
\label{tab:overall}
\centering
\setlength{\tabcolsep}{4pt}
\begin{tabular}{l|ccccc|cccccc}
\toprule
\multirow{2}{*}{Method} & \multicolumn{5}{c|}{Overall (\%)} & \multicolumn{6}{c}{Failure Rate per Category (\%)} \\
\cmidrule(lr){2-6} \cmidrule(lr){7-12}
 & B-F1 & RQ & FPCR & TPMR & $F_{1c}$ & Adware & Banker & Ransom. & Rootkit & Spyware & Trojan \\
\midrule
\framework & \textbf{69.39} & \textbf{56.21} & \textbf{96.00} & 93.13 & \textbf{45.70} & 70.83 & \textbf{37.29} & 36.84 & 100.00 & 16.33 & 53.85 \\
MalEval & 61.51 & 43.92 & 91.30 & 90.60 & 27.04 & 69.57 & 94.20 & 43.24 & 100.00 & 52.17 & 39.13 \\
LAMD & 57.20 & 39.93 & \textbf{96.00} & 78.26 & 28.63 & 73.91 & 90.00 & 68.42 & 100.00 & 43.90 & \textbf{33.33} \\
\midrule
\framework w/o domain grounding & 65.86 & 49.73 & \textbf{96.00} & 92.70 & 32.66 & 72.22 & 67.74 & 40.54 & 100.00 & 18.75 & 65.38 \\
\framework w/o semantics grounding & 64.36 & 49.22 & 92.00 & \textbf{98.28} & 35.16 & 82.61 & 76.81 & \textbf{29.73} & 100.00 & 24.44 & 58.33 \\
\framework w/o knowledge grounding & 63.25 & 54.03 & 88.00 & 91.30 & 39.74 & 75.00 & 63.79 & 37.84 & 100.00 & \textbf{9.09} & 65.22 \\
\framework w/o reviewer & 68.84 & 55.73 & 92.00 & 92.27 & 43.12 & \textbf{68.00} & 45.00 & 36.84 & 100.00 & 17.65 & 69.23 \\
\bottomrule
\end{tabular}
\end{table*}

% \begin{table}[t]
% \centering
% \caption{Report-level malware understanding results on MalEval.}
% \label{tab:overall}
% % \scriptsize
% \setlength{\tabcolsep}{2.2pt}
% % \renewcommand{\arraystretch}{1.08}

% \begin{tabular}{l|ccccc}
% \toprule
% Method & B-F1 & RQ & FPCR & TPMR & $F_{1c}$ \\
% \midrule
% \framework
% & \textbf{69.39}
% & \textbf{56.21}
% & \textbf{96.00}
% & 93.13
% & \textbf{45.70} \\

% MalEval
% & 61.51
% & 43.92
% & 91.30
% & 90.60
% & 27.04 \\

% LAMD
% & 57.20
% & 39.93
% & \textbf{96.00}
% & 78.26
% & 28.63 \\

% \midrule
% \framework\ w/o domain grounding
% & 65.86
% & 49.73
% & \textbf{96.00}
% & 92.70
% & 32.66 \\

% \framework\ w/o semantics grounding
% & 64.36
% & 49.22
% & 92.00
% & \textbf{98.28}
% & 35.16 \\

% \framework\ w/o knowledge grounding
% & 63.25
% & 54.03
% & 88.00
% & 91.30
% & 39.74 \\

% \framework\ w/o reviewer
% & 68.84
% & 55.73
% & 92.00
% & 92.27
% & 43.12 \\
% \bottomrule
% \end{tabular}
% \end{table}
\textbf{Structured Behavior Attribution.} 
We evaluate whether the system can translate its malware analysis into MITRE ATT\&CK Mobile technique attributions, which provides a standardized vocabulary for describing adversary behavior~\cite{xuan2026ttpdetect}. 
% A correct attribution requires not only identifying a suspicious behavior, but also mapping it to an appropriate standardized technique. 
% We use a 20-sample subset of malware sampled from MalEval for detailed ATT\&CK attribution analysis. We restrict this evaluation to samples from families for which reliable family-level ATT\&CK reference sets can be constructed from MITRE ATT\&CK pages, since other samples in the benchmark lack sufficiently detailed public technique-level ground truth.
% For each malware sample $a$ from family $f$, let $P(a) \subseteq \mathcal{T}$ denote the set of predicted MITRE ATT\&CK technique IDs, where $\mathcal{T}$ is the set of techniques applicable to mobile apps. 
ATT\&CK knowledge is commonly documented as a set of capabilities associated with a malware family aggregated across multiple variants and reports. Constructing reliable per-sample ATT\&CK ground truth would require expert reverse engineering of each APK, explicit technique-to-evidence traces, and adjudication of ambiguous cases. Such an annotation effort is beyond the scope of this system evaluation and constitutes a complementary benchmark-construction problem. We therefore construct a 20-APK attribution subset for which both family assignments and externally documented ATT\&CK references can be verified. For an APK $a$ assigned to family $f$, we use the documented technique set $G(f)$ as a reference for evaluating whether the generated attributions are consistent with known family capabilities. This setting provides a conservative and reproducible proxy for comparing systems under limited sample-level ground truth.
We extract $P(a)$ by parsing the technique IDs explicitly cited in the system's final report. 
% We use family-level rather than sample-level ground truth because per-sample technique annotation would require reverse-engineering effort comparable to the analysis task itself, which is not feasible at the scale of our benchmark. 
% This setting also matches the way operational threat-intelligence reports often describe TTPs at the family or campaign level~\cite{xuan2026ttpdetect}.
% Given $P(a)$ and $G(f(a))$, we define
% \begin{align}
% TP(a) &= |P(a)\cap G(f(a))|, \\
% FP(a) &= |P(a)\setminus G(f(a))|.
% \end{align}
Given $P(a)$ and $G(f(a))$, we define $TP(a)=|P(a)\cap G(f(a))|$ and $FP(a)=|P(a)\setminus G(f(a))|$.
We report macro-averaged precision and recall together with the aggregate counts sum-TP and sum-FP. 
For a fair comparison, we extend the analysis prompts of both MalEval and LAMD so that their final reports include MITRE ATT\&CK attribution.
We also canonicalize all predictions and reference labels against the official ATT\&CK documentation, mapping legacy identifiers to their current replacements, to avoid confounding the results with deprecated identifiers emitted by LLMs.

\textbf{Operational Reliability.} 
To measure self-consistency beyond one-shot accuracy, for each system and sample, we run the analysis three times as a cost-aware stability probe and compare the sets of predicted ATT\&CK techniques. We report Overall-Jaccard-Distance and TP-Jaccard-Distance. Overall-Jaccard-Distance measures the average pairwise distance between complete predicted technique sets:
\[
D_{\text{overall}}(P_i, P_j) =
1 - \frac{|P_i \cap P_j|}{|P_i \cup P_j|}.
\]
TP-Jaccard-Distance applies the same computation after intersecting predictions with the reference set, measuring whether correctly supported attributions remain stable across runs. Lower values indicate more stable predictions.
We also report semantic entropy~\cite{farquhar2024entropy} as a reference-free uncertainty signal over generated summaries.
% We also report semantic entropy~\cite{farquhar2024entropy} as a reference-free uncertainty signal over generated behavioral findings. For each sample, repeated report summaries are grouped together if their sentence-embedding cosine similarity exceeds a threshold $\tau=0.8$. This yields cluster counts $c_1,\dots,c_m$. We place a Dirichlet prior $\mathrm{Dir}(\alpha)$ over the class distribution, with $\alpha_k=0.5+c_k$, and report the posterior expected entropy:
% \[
% SE(a)=
% \mathbb{E}_{p \sim \mathrm{Dir}(\alpha)}
% \left[
% -\sum_k p_k \log p_k
% \right].
% \]
% which we estimate by Monte-Carlo sampling. Lower semantic entropy indicates that the system is more consistent across runs.

\subsubsection{Ablation Models}
% To isolate the contribution of each design choice in \framework, we construct five ablation variants, each disabling a single component while keeping the rest of the pipeline unchanged. We organize them along two axes that map to RQ2 and RQ3:
% \begin{itemize}
%     \item \emph{Grounding ablations} (RQ2): \framework w/o Semantics Grounding, w/o Domain Grounding, and w/o Knowledge Grounding (RAG).
%     \item \emph{Reliability ablations} (RQ3): \framework w/o Reviewer.
% \end{itemize}
To isolate the contribution of each design choice in \framework, we construct four ablation variants: (i) \textit{\framework w/o Domain Grounding} replaces the structured workflow with a single agent using the same semantic artifacts and retriever; (ii) \textit{\framework w/o Semantics Grounding} replaces static-analysis tools with generic navigation over JADX-decompiled files; (iii) \textit{\framework w/o Knowledge Grounding} removes MITRE ATT\&CK retrieval; and (iv) \textit{\framework w/o Reviewer} removes the final evidence-checking stage. 

\subsection{RQ1: Overall Effectiveness}

RQ1 evaluates whether \framework improves malware understanding compared with existing LLM-based malware-analysis methods and frontier agentic systems. We analyze the results along two complementary dimensions. Table~\ref{tab:overall} measures report-level malware understanding, and Table~\ref{tab:attack} measures structured behavior attribution. 
% Together, the two tables show not only whether a system detects malware, but also whether it produces useful, evidence-supported behavioral explanations.

\textbf{Comparison with prior LLM-based malware-analysis pipelines.}
Compared with MalEval and LAMD, \framework achieves the strongest report-level performance across the main metrics. 
% Relative to MalEval, \framework improves B-F1 and RQ by 12.8\% and 28.0\%, respectively. Relative to LAMD, the improvements are larger, with B-F1 and RQ increasing by 21.3\% and 40.8\%. 
% The category-level F1 also improves substantially over both systems, suggesting that \framework is better at producing behaviorally meaningful analyses across diverse malware families rather than only improving the final binary verdict. 
The category-level F1 also improves substantially over both systems, indicating better aggregate discrimination across malware categories rather than only an improved final binary verdict. The per-category failure rates in Table~I reveal a more heterogeneous picture. \framework achieves the lowest failure rates on Banker, Ransomware, and Spyware, but all methods fail on Rootkit samples and \framework remains weaker on Trojans. This indicates that grounded reasoning improves overall behavior understanding while some malware categories remain challenging.
MalEval relies on single-pass reasoning over provided context, while LAMD extracts context around sensitive APIs. Their weaker performance indicates that suspicious code or static facts are not sufficient for reliable malware understanding. \framework improves over these systems because it structures the task as an iterative reasoning process and explores regions adaptively. The ATT\&CK results in Table~\ref{tab:attack} further support this conclusion. Compared with MalEval and LAMD, \framework achieves higher ATT\&CK coverage and precision while reducing false-positive attributions by 57.0\% and 35.1\%, respectively. 
% These results show that \framework also improves structured behavior attribution, where the system must decide which observed program behaviors justify specific ATT\&CK technique labels.

\textbf{Comparison with frontier agentic systems.}
Frontier agentic systems exhibit a different strength. Claude Code (Opus-4.7) achieves the highest recall and the largest number of true-positive technique attributions, but produces substantially more false-positive attributions. In comparison, \framework reduces false positives by 68.9\% and achieves higher Macro-Precision. This contrast shows that \framework occupies a different operating point from frontier agents. Claude Code explores broadly and surfaces many plausible behaviors, but this open-ended exploration can over-map real low-level observations to unsupported ATT\&CK techniques. \framework is more reliability-oriented: it reports fewer technique IDs, but the reported attributions are more tightly constrained by localized evidence and retrieved threat knowledge. For malware understanding, this distinction matters because analysts must be able to audit why a behavior was claimed. A higher-recall system is useful for broad triage, but a lower-false-positive system is often more suitable when behavioral claims must be trusted, reviewed, or integrated into downstream intelligence workflows. The comparison with Codex shows a similar but less extreme pattern. Codex (GPT-5.4) achieves competitive precision, but lower ATT\&CK recall than \framework. \framework improves Macro-Recall and Macro-Precision over Codex (GPT-5.4) while reducing cost by 51.0\%. This suggests that a grounded malware-analysis harness can provide a more favorable capability--cost balance than a general-purpose coding-agent harness.

\textbf{Harness design matters beyond base-model capability.}
The model-controlled frontier-agent variants further isolate the effect of the harness. When the same DeepSeek-V3.2 model is placed inside generic Codex or Claude Code harnesses, \framework achieves much higher attribution precision with far fewer false positives. This indicates that the performance gap cannot be explained by base-model capability alone. The surrounding harness, including domain-specific reasoning roles, semantic evidence localization, review, and knowledge-grounded attribution, is central to reliable malware understanding. Overall, \framework improves over existing LLM-based malware-analysis pipelines in both report-level understanding and structured behavior attribution. Compared with frontier agentic systems, it provides a more conservative and cost-efficient operating point with stronger attribution precision and substantially fewer unsupported claims. 
% These results support the central thesis of this work: reliable malware understanding requires not only a capable model, but also a grounding-aware reasoning harness that controls how hypotheses are generated, how evidence is localized, and how behavior labels are attributed.

% \input{Table/Attack}

% \begin{figure}[t]
%   \centering
%   \includegraphics[width=\columnwidth]{Figure/precision_heatmap.pdf}
%   \caption{Per-sample MITRE ATT\&CK \emph{precision} over the same samples and systems.}
%   \label{fig:precision_heatmap}
% \end{figure}

\subsection{RQ2: Grounded Reasoning Analysis}
RQ2 examines whether the three grounding mechanisms are necessary for reliable malware reasoning. We interpret each ablation through the role it plays in the reasoning process.
% domain grounding shapes hypothesis generation, semantics grounding supports evidence localization, and knowledge grounding constrains behavior attribution. 
% This distinction matters because malware understanding is not a single-stage prediction problem. A system may fail by forming the wrong hypothesis, inspecting the wrong evidence, or mapping a valid behavior to an unsupported ATT\&CK technique. 
% The ablations show that each grounding mechanism reduces a different class of errors.

\textbf{Domain grounding controls hypothesis generation.} Removing domain grounding degrades both report-level quality and ATT\&CK coverage. As shown in Table~\ref{tab:overall}, \framework w/o domain grounding reduces B-F1, RQ, and category-level F1 by 5.1\%, 11.5\%, and 28.5\%, respectively, relative to the full system. The ATT\&CK results in Table~\ref{tab:attack} show the same trend: Macro-Recall decreases by 24.1\%, and the number of true-positive technique attributions decreases by 41.3\%. The failure lies in how hypotheses are formed and revised. Without the analyst-inspired workflow, the model is more likely to stop at shallow observations, merge unrelated signals into broad summaries, or fail to follow a suspicious lead until it becomes a behavior-level claim. The lower number of false positives in the ATT\&CK table should therefore not be read as improved reliability. It mostly reflects under-generation: the agent makes fewer claims and misses many valid behaviors. 

\begin{table*}[t]
\centering
\caption{Structured behavior attribution results on the 20-sample subset, grouped by capability, reliability, stability, and operational cost per sample. Capability and reliability are averaged over three runs, where $\pm$ denotes standard deviation.}
\label{tab:attack}
\scriptsize
\setlength{\tabcolsep}{2pt}
\renewcommand{\arraystretch}{1.05}
\begin{tabular}{l|cc|cc|ccc|cc}
\toprule
\multirow{2}{*}{Method}
& \multicolumn{2}{c|}{Capability}
& \multicolumn{2}{c|}{Reliability}
& \multicolumn{3}{c|}{Stability}
& \multicolumn{2}{c}{Operational Cost} \\
\cmidrule(lr){2-3}
\cmidrule(lr){4-5}
\cmidrule(lr){6-8}
\cmidrule(lr){9-10}
& Macro-Recall & sum-TP
& Macro-Precision & sum-FP
& Overall-Jaccard-Distance & TP-Jaccard-Distance & Semantic Entropy
& Cost & Tokens \\
\midrule
\textbf{Malaika}
& $24.56_{\pm1.46}$ & $75_{\pm5}$
& $\mathbf{67.61}_{\pm0.84}$ & $37_{\pm2}$
& \textbf{39.22} & 34.96 & 0.12
& \$0.33 & 1.32M \\

\quad w/o domain grounding
& $18.64_{\pm1.51}$ & $44_{\pm4}$
& $66.77_{\pm3.98}$ & $\mathbf{22}_{\pm1}$
& 47.25 & 40.83 & 0.31
& \$0.13 & 0.43M \\

\quad w/o semantics grounding
& $13.23_{\pm1.86}$ & $42_{\pm6}$
& $27.88_{\pm3.23}$ & $121_{\pm13}$
& 83.18 & 70.75 & 0.33
& \textbf{\$0.12} & \textbf{0.41M} \\

\quad w/o knowledge grounding
& $15.99_{\pm0.73}$ & $45_{\pm1}$
& $31.68_{\pm3.22}$ & $97_{\pm7}$
& 71.47 & 71.32 & 0.15
& \$0.19 & 0.75M \\

\quad w/o reviewer
& $25.64_{\pm0.94}$ & $78_{\pm2}$
& $64.99_{\pm0.14}$ & $40_{\pm5}$
& 40.45 & 30.59 & 0.12
& \$0.17 & 0.70M \\

\quad w/ GPT-5.4
& $22.84_{\pm1.32}$ & $64_{\pm5}$
& $42.10_{\pm2.18}$ & $57_{\pm12}$
& 50.27 & 50.22 & 0.11
& \$1.67 & 1.19M \\

\midrule
Claude Code (Opus-4.7)
& $51.30_{\pm0.95}$ & $\mathbf{159}_{\pm2}$
& $58.99_{\pm0.29}$ & $110_{\pm1}$
& 41.88 & \textbf{27.24} & 0.14
& \$1.54 & 1.42M \\

Claude Code (Opus-4.7 + skill)
& $\mathbf{51.45}_{\pm1.13}$ & $157_{\pm4}$
& $57.31_{\pm4.20}$ & $119_{\pm15}$
& 47.72 & 32.81 & 0.14
& \$1.70 & 1.80M \\

Claude Code (DeepSeek-V3.2 + skill)
& $12.70_{\pm2.27}$ & $41_{\pm4}$
& $15.74_{\pm3.56}$ & $209_{\pm9}$
& 90.19 & 86.43 & 0.17
& \$0.42 & 1.52M \\

Codex (GPT-5.4)
& $18.78_{\pm5.59}$ & $56_{\pm13}$
& $56.18_{\pm2.35}$ & $41_{\pm2}$
& 79.71 & 77.67 & \textbf{0.03}
& \$0.96 & 1.41M \\

Codex (DeepSeek-V3.2)
& $14.32_{\pm0.80}$ & $46_{\pm2}$
& $19.74_{\pm1.39}$ & $171_{\pm3}$
& 90.51 & 76.84 & 0.30
& \$0.34 & 1.20M \\

MalEval
& $11.77_{\pm1.64}$ & $37_{\pm5}$
& $32.56_{\pm8.63}$ & $86_{\pm29}$
& 86.03 & 68.83 & 0.42
& \$0.37 & 1.32M \\

LAMD
& $5.54_{\pm1.72}$ & $13_{\pm3}$
& $14.31_{\pm3.01}$ & $57_{\pm2}$
& 82.52 & 94.44 & 0.30
& \$0.23 & 0.83M \\
\bottomrule
\end{tabular}
\end{table*}
\textbf{Semantics grounding controls evidence localization.} Similar to \framework w/o domain grounding, removing semantics grounding results in decreases in B-F1, RQ, and category-level. Notably, it causes the largest reliability loss in ATT\&CK attribution, and the number of false-positive attributions increases from 37 to 121, indicating that the system becomes much more prone to unsupported mappings. 
% The agent still follows an analyst-style workflow, but it reasons over poorly targeted context. When it must explore raw decompiled files with generic file-system tools,
The model tends to rely on surface signals such as permissions, suspicious strings, or isolated functions. These signals are useful starting points, but they do not establish behavior on their own. Without semantic localization, the agent often fails to connect dispersed evidence, producing both missed behaviors and unsupported claims. 
% Semantics grounding is therefore necessary because it determines which program facts are exposed to the reasoning process and whether they are presented in a form that supports behavior-level inference.

\textbf{Knowledge grounding controls behavior attribution.} Without MITRE retrieval, report-level RQ decreases by only 3.9\%, suggesting that domain and semantics grounding still allow the system to identify many code-level behaviors. ATT\&CK attribution, however, becomes much less calibrated: Macro-Recall decreases by 34.9\%, Macro-Precision decreases by 53.1\%, and false positives increase by 162.2\%. This result shows that evidence localization alone is not enough for standardized behavioral attribution. Without external threat knowledge, the model relies on parametric knowledge from LLMs when assigning technique IDs. This leads to several attribution errors, such as mapping a real behavior to an overly broad technique, selecting a stronger technique than the code supports, or using outdated or invalid identifiers. 
% Knowledge grounding constrains how that behavior is named and mapped to an externally verifiable taxonomy. Its main contribution is not to find additional code evidence, but to calibrate the final behavior labels.

% Overall, the ablation results support the tri-grounded mechanisms of malware understanding. 
% Removing domain grounding weakens hypothesis generation and leads to incomplete behavioral coverage. Removing semantics grounding disrupts evidence localization and makes the model reason from poorly targeted context, causing both missed behaviors and unsupported claims. Removing knowledge grounding preserves much of the code-level analysis but makes ATT\&CK attribution substantially less reliable.
These complementary failure modes show that the proposed groundings correspond to distinct reasoning functions required for reliable malware understanding: forming plausible hypotheses, locating and exploring relevant evidence, and assigning calibrated behavioral labels.

% \subsection{RQ3: Reliability and Cost Analysis}
\subsection{RQ3: Trustworthiness and Practicality Analysis}
RQ3 examines whether \framework produces trustworthy malware analyses in practice. We assess three complementary properties. First, we evaluate whether the Reviewer improves evidence discipline in final reports. We then assess output stability across repeated executions. Finally, we analyze whether these benefits are achieved at a practical inference cost.

\subsubsection{Reviewer Ablation}
Removing the Reviewer slightly affects report-level metrics, while it has a clearer effect on attribution discipline. In Table~\ref{tab:attack}, removing the Reviewer slightly increases Macro-Recall and true-positive attributions, but decreases Macro-Precision and increases false positives. This pattern suggests that the Reviewer acts as a conservative filter: it filters borderline mappings after evidence has been localized and attributed. Rather than discovering new evidence, it reduces residual unsupported claims in the final report.
% Without it, more borderline technique mappings are allowed into the final report, which can recover a small number of additional true positives but also introduces more unsupported claims. With the Reviewer, \framework sacrifices a small amount of coverage for stronger precision and fewer false-positive attributions. This result clarifies the role of the Reviewer. It is not the main source of new evidence, nor is it expected to dramatically change recall. Instead, it checks whether a proposed behavior claim is sufficiently supported, whether the reasoning connects the relevant evidence, and whether the ATT\&CK attribution follows from the observed behavior. The relatively small but consistent precision gain is therefore expected: the Reviewer primarily corrects residual reasoning errors after the grounding mechanisms have already constrained exploration and attribution.

\subsubsection{Self-Consistency}
A system that gives substantially different analysis for the same sample across different runs is difficult to trust. 
As shown in Table~\ref{tab:attack}, \framework achieves lower Overall-Jaccard-Distance than the prior LLM-based malware-analysis pipelines and the frontier agentic baselines. 
% Compared with Claude Code with the Android malware skill, \framework reduces overall prediction-set instability by 17.8\%. Compared with MalEval and LAMD, the reduction is more substantial, at 54.4\% and 52.5\%, respectively. 
This indicates that \framework is less likely to produce highly variable ATT\&CK attribution sets across repeated runs. The TP-Jaccard-Distance results show a more nuanced picture. \framework is more stable than MalEval, LAMD, and Codex (GPT-5.4) on correctly supported attributions, but is slightly less stable than Claude Code (Opus-4.7+skill) on true-positive predictions. This is consistent with the different operating points of the systems. Claude Code (Opus-4.7+skill) explores broadly and repeatedly recovers many true-positive techniques, but it also produces more false positives. \framework is more conservative: it reports fewer techniques overall, and its prediction is more stable and precise. Semantic entropy provides a complementary view. \framework has substantially lower semantic entropy than Claude Code (Opus-4.7+skill), MalEval, and LAMD, suggesting that its generated summaries are more semantically consistent across repeated executions. 
Codex (GPT-5.4) has lower semantic entropy, but also lower ATT\&CK recall and lower precision than \framework. This illustrates a limitation of stability signals alone: a system can appear stable if it consistently under-reports behavior. For this reason, we interpret self-consistency together with attribution precision and recall rather than as a standalone quality metric.

\subsubsection{Operational Cost}
\textsf{Malaika}'s additional cost supports capabilities that the ablations lack: removing a grounding component lowers the budget but weakens capability, while removing the Reviewer increases false-positive attributions. 
Despite this overhead, \framework remains economical compared with all baselines except LAMD. Relative to frontier agentic systems and models, it achieves substantially lower cost while maintaining stronger attribution precision and fewer false positives. These results indicate that the complete grounded harness provides a favorable capability--cost trade-off.

Overall, \framework offers a practical reliability-oriented operating point for malware understanding: it is not designed to maximize the number of reported behaviors, but to ensure that reported behaviors are stable and reliable.

\section{Case Study}

% Muted, print-friendly colors
\definecolor{FPInvBg}{RGB}{252,232,230}
\definecolor{FPInvFg}{RGB}{150,42,35}

\definecolor{FPMechBg}{RGB}{254,241,224}
\definecolor{FPMechFg}{RGB}{145,86,22}

\definecolor{FPPermBg}{RGB}{239,235,249}
\definecolor{FPPermFg}{RGB}{92,70,145}

\definecolor{FPGTBg}{RGB}{226,240,246}
\definecolor{FPGTFg}{RGB}{40,104,130}

\definecolor{FPBlBg}{RGB}{238,238,238}
\definecolor{FPBlFg}{RGB}{90,90,90}

\newcommand{\fpbadge}[3]{%
  \begingroup
  \setlength{\fboxsep}{1.2pt}%
  \colorbox{#2}{\textcolor{#3}{\scriptsize\bfseries #1}}%
  \endgroup
}

\newcommand{\taginv}{\fpbadge{INV}{FPInvBg}{FPInvFg}}
\newcommand{\tagmech}{\fpbadge{MECH}{FPMechBg}{FPMechFg}}
\newcommand{\tagperm}{\fpbadge{PERM}{FPPermBg}{FPPermFg}}
\newcommand{\taggtgap}{\fpbadge{GT-GAP}{FPGTBg}{FPGTFg}}
\newcommand{\tagbl}{\fpbadge{BL}{FPBlBg}{FPBlFg}}

\providecommand{\evfp}[3]{%
#1\par
#2\hspace{2pt}#3%
}
\setlength{\textfloatsep}{6pt plus 1pt minus 2pt}
\begin{table}[t]
\centering
\caption{False-positive ATT\&CK attributions in the GolfSpy case study. Claude Code reports 9 false positives, and \framework reports 3 false positives.}
\label{tab:casestudy_fp}
\scriptsize
\setlength{\tabcolsep}{3pt}
\renewcommand{\arraystretch}{1.06}
\begin{tabular}{@{}p{0.11\columnwidth}@{\hspace{4pt}}|@{\hspace{5pt}}p{0.85\columnwidth}@{}}
\toprule
Technique & Evidence and FP reason \\
\midrule

\multicolumn{2}{@{}l}{\textbf{Claude-Code-Opus-4.7 (skill)} --- 9 FP} \\
\midrule

T1437.002 &
\evfp{Socket ``Signal'' line protocol.}
{\taginv}
{This Mobile ATT\&CK sub-technique does not exist.} \\
\midrule

T1521.001 &
\evfp{C2 strings are Base64/XOR-transformed.}
{\tagmech}
{Lightweight obfuscation fits T1406, not an encrypted channel.} \\
\midrule

T1541 &
\evfp{\texttt{GlassService} is restarted by a 1200\,s watchdog.}
{\tagmech}
{Not foreground-service persistence.} \\
\midrule

T1626.001 &
\evfp{Requests SMS, contacts, call-log, camera, location, and overlay permissions.}
{\tagmech}
{Dangerous permissions do not imply device-admin elevation.} \\
\midrule

T1644 &
\evfp{Media files are staged under hidden \texttt{/ecap32x} paths.}
{\tagmech}
{This is local cache staging, not an out-of-band channel.} \\
\midrule

T1577 &
\evfp{A benign prayer app embeds the \texttt{c204} spyware payload.}
{\tagbl}
{This technique targets modifying another app's executable.} \\
\midrule

T1422 &
\evfp{Reads \texttt{ConnectivityManager.getNetworkInfo} before upload.}
{\taggtgap}
{Real network-state use, but absent from the family-level GT.} \\
\midrule

T1437.001 &
\evfp{Uploads stolen files via \texttt{multipart/form-data} POSTs.}
{\taggtgap}
{Valid C2 transport, but GT credits as T1646 exfiltration.} \\
\midrule

T1655.001 &
\evfp{Masquerades as a Sunni prayer/qibla utility.}
{\taggtgap}
{Real masquerading, but absent from the family-level GT.} \\

\specialrule{0.8pt}{2pt}{2pt}

\multicolumn{2}{@{}l}{\textbf{\framework} --- 3 FP} \\
\midrule

T1417.002 &
\evfp{Requests \texttt{SYSTEM\_ALERT\_WINDOW}.}
{\tagperm}
{Overlay capability is inferred from permission only; no GUI-input-capture code is observed.} \\
\midrule

T1623 &
\evfp{Socket ``Signal'' dispatcher selects in-app actions.}
{\tagmech}
{No \texttt{Runtime.exec} \texttt{ProcessBuilder} or shell strings found.} \\
\midrule

T1603 &
\evfp{Uses \texttt{AlarmManager} for repeating background services.}
{\taggtgap}
{Valid scheduling behavior, but absent from the family-level GT.} \\

\midrule
\multicolumn{2}{@{}p{\columnwidth}@{}}{\scriptsize
\taginv\ invalid ATT\&CK identifier;\;
\tagmech\ evidence supports a weaker or different mechanism;\;
\tagperm\ inferred from permission alone;\;
\taggtgap\ valid behavior missing from ground truth;\;
\tagbl\ borderline attribution.} \\
\bottomrule
\end{tabular}
\end{table}

We compare \framework with Claude Code (Opus 4.7+skill) on a representative sample from the GolfSpy family. We inspect both the final reports and the reasoning traces, and manually verify each attribution against the sample. Both systems recover the main surveillance behaviors, including device fingerprinting, SMS and call monitoring, C2 exfiltration, and persistence. They reach these findings through different workflows. Claude Code conducts broad, open-ended exploration over the APK workspace, whereas \framework starts from suspicious seeds, reconstructs localized context, retrieves relevant ATT\&CK knowledge, and applies a review stage before reporting.
The main difference lies in attribution calibration. 
Claude Code reports 8 true positives and 9 false positives, whereas \framework reports the same number of true positives with 3 false positives. 
As shown in Table~\ref{tab:casestudy_fp}, most Claude Code false positives are over-attributions from real low-level evidence. For example, it maps lightweight obfuscation to an encrypted channel, interprets dangerous permissions as device-administrator elevation, and treats local cache staging as an out-of-band channel. 
% Manual inspection confirms that the sample contains no cryptographic cipher, declares no device-administrator component, and invokes no shell-execution API.
% The excluded false positives reflect the granularity of the family-level ground truth rather than fabricated evidence.
\framework also makes over-attributions, 
% including inferring GUI input capture from an overlay permission and command execution from an in-app socket dispatcher. 
but the grounding mechanisms reduce such errors. Some false positives in both systems are code-grounded behaviors that are absent from the family-level reference set, reflecting the granularity of the ground truth rather than fabricated reasoning. Overall, the case study shows that the two systems differ less in raw evidence discovery than in how conservatively they convert evidence into ATT\&CK claims. Claude Code surfaces rich artifacts through broad exploration, while \framework produces tighter and more directly auditable attributions.
\section{Discussion}

\textbf{Capability--Cost--Sovereignty Trade-offs.}
Malware analysis agents should be evaluated not only by model capability in isolation. Closed frontier models offer strong reasoning and code-understanding ability, but malware analysis frequently involves sensitive binaries, proprietary applications, and incident-response artifacts. On the other hand, open-weight models can be deployed in controlled environments, fine-tuned for specialized tasks, and examined more directly for interpretability and uncertainty. \framework helps narrow this gap, enabling conservative and auditable malware understanding at lower cost.

% \subsection{Extension on dynamic analysis and native code analysis}
% \framework is intentionally designed around static evidence, which provides scalability and auditability because every reported behavior can be traced back to concrete artifacts. However, static analysis cannot fully observe behaviors that depend on runtime activation, dynamic code loading, native code, or external infrastructure. These limitations clarify the scope of \framework rather than weakening it: the framework provides an evidence-grounded static reasoning harness for behaviors visible from program artifacts. A natural extension is to incorporate dynamic analysis and native-code analysis into the same hypothesis-and-review loop, so that runtime evidence can validate hypotheses that static evidence only partially supports.

\textbf{Beyond Android Malware Understanding.}
% Although this work instantiates \framework for Android, the framework is better understood as a general harness for grounded code analysis under partial observability. The three forms of grounding are platform-independent. Domain grounding defines a role-specialized workflow. Semantics grounding requires program evidence to be supplied through static-analysis tools. Knowledge grounding retrieves from an external inspectable taxonomy and supports the claim. This makes transfer largely a matter of retargeting the tool and knowledge layers. More broadly, the harness is not limited to malware analysis. The same problem of reconstructing behavior from partial and dispersed program evidence, and verifying each claim against that evidence, also appears in vulnerability triage, repository-level code auditing, and supply-chain risk assessment. The framework can therefore be viewed as a grounded-reasoning architecture for software analysis, with Android malware understanding as one instantiation.
Although \framework is instantiated for Android malware understanding, its broader contribution concerns evidence-grounded software reasoning under partial observability. Across many software-analysis settings, analysts must reconstruct high-level behaviors or security implications from dispersed evidence while determining whether the available evidence is sufficient to support the claimed conclusion. Malware understanding is one instance of this general reasoning problem. The underlying goal remains the same: producing conclusions whose scope and confidence are justified by the available evidence.

\textbf{Grounding-Aware Agents and Reproducible Security Evaluation.}
More broadly, grounding-aware agent design may improve the reproducibility of security-agent evaluations. Security-agent performance is often affected by prompt wording, tool instructions, and manually written skills, but these factors are often treated as implementation details rather than controlled experimental variables. By embedding reasoning constraints into explicit grounding mechanisms, \framework shifts part of the analytical burden from prompt engineering toward structured and inspectable workflows, making future comparisons more reproducible and easier to interpret.
\section{Threats to Validity}
\label{sec:threats}

\textbf{Construct validity.} 
Our evaluation uses report-level scores, structured ATT\&CK attribution, stability, and operational cost as proxies for malware understanding. These metrics do not capture every aspect of analyst utility. Report-level scores rely partly on LLM-as-a-judge evaluation, while uncertainty metrics such as Jaccard distance and semantic entropy measure stability rather than factual correctness. The ATT\&CK evaluation is also limited by its 20-sample subset and family-level ground truth, since reliable technique-level references are only available for limited samples. As a result, some sample-specific behaviors may be counted as false positives if they are absent from the family-level ground truth. We therefore interpret ATT\&CK results as a proxy for structured behavior reasoning rather than complete per-sample ground truth.

\textbf{Internal validity.} 
% Although prompts are fixed across all experiments, prompt sensitivity remains an inherent limitation of LLM-based systems. Differet prompt formulations may affect absolute performance estimates. Our goal is not to optimize prompts, but rather to evaluate the impact of grounding mechanisms under a fixed and reproducible prompting configuration.
LLM-based systems are sensitive to prompts, model versions, tool interfaces, and harness implementations. Baseline comparisons may also be affected by the partially closed nature of frontier systems. We mitigate these threats by reporting model versions, reasoning settings, and skill configurations, and by adding DeepSeek-controlled Codex and Claude Code baselines to better isolate harness effects from base-model capability. \framework also depends on static-analysis artifacts such as decompiled code, call graphs, sensitive APIs, and suspicious strings, whose imprecision may affect the evidence exposed to agents.

\textbf{External validity.}
Our evaluation focuses on Android and the MalEval benchmark. Although this setting is appropriate for behavior-level malware understanding, the benchmark size and Android-only scope limit generalization to other malware ecosystems. The current framework is also intentionally static: it can reason about behaviors visible from program artifacts, but may miss behaviors that depend on runtime activation, dynamic code loading, or native code. While the grounding principles are not Android-specific, applying \framework to other platforms would require retargeting the semantic-analysis tools and external knowledge base. Therefore, the absolute numbers should be interpreted as benchmark- and version-specific, while the main conclusion concerns the comparative effect of grounding-aware harness design under the evaluated setting.

\section{Related Work}
% This section reviews how Android malware detection has shifted from conventional learning-based approaches to LLM-powered approaches, while also introducing recent advances in agentic AI for software security and situating our work in this evolving field.

\subsection{From Learning-based to LLM-powered Malware Analysis}
Machine learning-based approaches~\cite{arp2014drebin, deepdrebin2017, he2022msdroid} have been widely adopted to improve scalability and enable automated classification at inference time. However, their effectiveness remains fragile under temporal drift~\cite{kan2024tesseract, pendlebury2019tesseract, aurora}, requiring continuous human supervision and periodic retraining as threats evolve~\cite{continuous2023, zheng2025tif}. Although prior works have explored feature-level explanations, these explanations are typically model-centric and fail to provide actionable human-readable insights to security analysts~\cite{he2023finer}. This lack of interpretability limits their practical utility in high-stakes security environments. More importantly, most methods are optimized for binary or family classification, whereas practical malware understanding requires behavior-level reasoning: identifying malicious behaviors and their related evidence. 

Recent work has begun to use LLMs to bridge low-level behavioral observations and malware detection. For example, Trident~\cite{saul2026trident} uses LLMs to convert sandbox behavior reports into behavior-based detection rules for drift-resilient malware detection. For malicious-code reasoning, however, the challenge goes beyond recognizing behavioral indicators: it requires explaining how concrete program operations and their execution context support a malicious-behavior claim. LLMs offer a natural interface for connecting dispersed program observations into higher-level behavioral interpretations. Yet, raw whole-program prompting does not by itself ensure reliable reasoning. Although some modern models can process long input contexts, semantic reasoning over long contexts remains fragile~\cite{suman2025sense}, particularly when relevant evidence is sparse, distributed across a codebase, and intertwined with benign functionality. This has motivated structured context selection and hierarchical analysis. MalParse~\cite{walton2024malparse}, for instance, analyzes programs across function, class, and package levels. LAMD~\cite{qian2025lamd} further shows that sensitive-context extraction and tier-wise code reasoning can improve robustness under temporal drift in malware classification. Complementarily, MalEval~\cite{zheng2025maleval} introduced a principled framework for benchmarking LLM reasoning and reliability, offering the first reproducible foundation for assessing semantic alignment. These advances improve context management and code reasoning, but still rely on fixed analysis workflows, making them brittle when malicious evidence is fragmented across multiple components and implicit execution paths. 
% Our work builds on these advances, but moves beyond static prompting pipelines by explicitly studying how grounded, iterative reasoning can further enhance analytical quality.

\subsection{Agentic AI for Security-centric Analysis} 
LLM-based agents extend single-pass analysis pipelines by combining model reasoning with task decomposition, task-specific tools, external information, and memory. These capabilities enable agents to inspect content on demand and iteratively update analysis. Recent work in vulnerability detection, code localization, and program repair~\cite{guo2025repoaudit, chen2025locagent, yu2025patchagent, li2025patchpilot} demonstrates the potential of this paradigm for security-centric code analysis. However, reasoning and tool use alone do not guarantee reliable conclusions: a final claim is auditable only when it can be traced to sufficient and checkable evidence. 

Despite this progress, the role of agentic AI in malware detection and analysis remains underexplored. Existing malware detection works have begun to adopt agent-based frameworks~\cite{masdroid, zeng2026mard, androsem}, but their evaluations remain largely classification-centric, focusing on whether the predicted label is correct rather than whether the agent faithfully reconstructs the underlying malicious behaviors. This distinction is crucial because an agent may produce a correct detection result while relying on incomplete or irrelevant evidence.

Recent frontier industrial systems, such as OpenAI's Codex~\cite{openai2026codexcybersafety} and Anthropic's Claude Code~\cite{anthropic2026claudecodesecurity}, demonstrate remarkable capability in code reasoning, especially around vulnerability discovery and validation~\cite{carlini2026mythos, zheng2026veritas}. Agent skills~\cite{anthropic2025skills, li2026skillsbench} further empower agents with domain-specific expertise by combining task-specific instructions during problem solving. While these mechanisms can encode analyst expertise, they do not by themselves ensure that claims are supported by sufficient evidence.

% Given these advances, the key question is how this capability transfers to a different but equally important task: malicious-code understanding. Unlike vulnerability discovery, malware understanding requires reconstructing dispersed behaviors, linking them to concrete program evidence, and deciding whether a malicious-behavior claim is sufficiently supported. 
% More importantly, can we design a system with open-weight models that narrows the gap to frontier systems when wrapped in appropriate harness while retaining predictable cost and stronger data-governance control?
\section{Conclusion}

% This paper studies malware understanding as a grounded reasoning problem. Behavior-level analysis requires reconstructing how dispersed program fragments interact and verifying whether the conclusion is supported by concrete code evidence. We present \framework{}, a tri-grounded multi-agent framework that combines static analysis, analyst-inspired review, and external threat-knowledge for reliable malware understanding. Our evaluation shows that \framework{} improves over existing LLM-based malware-analysis pipelines in both analysis quality and behavior attribution. Compared with frontier agentic systems, \framework{} offers a more conservative, precision-oriented, and lower-cost alternative. More broadly, this work highlights a key direction for future security agents: reliable malware analysis should be built around evidence construction, claim verification, and auditable reasoning.

We formulate malware understanding as grounded reasoning under partial observability and present \framework, a tri-grounded framework for reconstructing auditable behaviors from dispersed program evidence. \framework combines domain grounding for analyst-inspired hypothesis generation and review, semantics grounding for execution-relevant evidence localization, and knowledge grounding for verifiable ATT\&CK attribution. Our results show that this design improves report quality and attribution precision over prior LLM-based malware-analysis pipelines, while providing a more conservative and lower-cost alternative to frontier agentic systems. The ablations further show that the three grounding mechanisms address complementary needs in hypothesis formation, evidence connection, and behavioral attribution. Overall, reliable malware analysis depends not only on model capability, but also on how reasoning is grounded in evidence.

\bibliographystyle{IEEEtran}
\bibliography{reference}
\appendices
% \input{Tex/ExpSetting}
% \input{Tex/Baseline}
% \input{Tex/Dataset}
% \input{Tex/Prompt}
% \input{Tex/Algorithm}

% % \bibitem{IEEEhowto:kopka}

% % H.~Kopka and P.~W. Daly, \emph{A Guide to \LaTeX}, 3rd~ed.\hskip 1em plus
% %   0.5em minus 0.4em\relax Harlow, England: Addison-Wesley, 1999.
% \end{thebibliography}

\end{document}